# Music therapy for improving mental health in offenders: protocol for a systematic review and meta-analysis

XiJing Chen[1], Helen Leith[1], Leif Edvard Aarø[2], Terje Manger[3], Christian Gold[4]

1. Department of Communication and Psychology, Aalborg University, Aalborg, Denmark.
2. Division of Mental Health, Norwegian Institute of Public Health, Oslo/Bergen, Norway.
3. Department of Psychosocial Science, University of Bergen, Norway.
4. Grieg Academy Music Therapy Research Centre (GAMUT), Uni Health, Uni Research, Bergen, Norway

Contact address: Christian Gold, GAMUT – The Grieg Academy Music Therapy Research Centre, Uni Research Health, Uni Research, Lars Hilles gate 3, Bergen, 5015, Norway. christian.gold@uni.no

# Abstract

This is a protocol for a systematic review of the effects of music therapy on offenders. Based on randomised controlled trials, the review aims to assess the effectiveness of music therapy on adolescent and adult offenders in custodial institutions including forensic psychiatric hospitals, and offenders or probationers in the community. The outcomes to be evaluated include alleviated symptoms of mental illness, psychosocial competencies and reduced recidivism.

Note: This protocol was developed with and for the Cochrane Collaboration in 2012-2013, but was not published in the Cochrane Database of Systematic Reviews. Co-author H.L. died in 2014 and the remaining authors would like to acknowledge her contribution. She did approve the current version of the text in 2013.

# BACKGROUND

## Description of the Condition

An offender is a person who has been convicted of a criminal offence or who has been adjudged to have committed an act of juvenile delinquency. More than 10.65 million offenders were held in penal institutions throughout the world in 2008 (Walmsley 2009). Criminal justice systems usually deal with juvenile offenders differently to adult offenders, and the age of criminal responsibility varies greatly. The United Nations (UN) considers criminal responsibility below the age of 12 years unacceptable, but not all countries adhere to that and the age of criminal responsibility ranges from six to 18 years across countries (Prison Reform Trust 2011, p34).

The prevalence of mental health problems is high in the offending population (James 2006; WHO 2007). Although a small proportion of offenders with a diagnosed mental illness are treatedin specialist forensic psychiatric hospitals, most offenders with mental health problems are serving custodial or community sentences. Many such mental health problems will be either undiagnosed or will not fully meet the criteria of a mental disorder. However,high prevalence rates of mental disorders are also likely to signify a high prevalence of other mental health problems. For example, up to 90% of offenders in prison in the UK (95% in the case of young offenders) have a diagnosis of mental illness or a drug and/oralcohol dependency-related mental or behavioural disorder (Prison Reform Trust 2011). In China, 49% to 71% of male offenders have a personality disorder (Wang 2007). In the US, 56% of offenders in prison have symptoms of mental disorder or a history of treatment for mental disorder, 43% have symptoms or a history of mania, 23% have a history of major depression and 15% have experienced psychotic symptoms such as hallucinations or delusions (James 2006). They often show paranoia, suicidal ideation, self-isolating and self-harming behaviour (Frühwald 2005). Many prisoners also have symptoms that resemble the negative symptoms of schizophrenia -blunted affect, low motivation and poor social relationships – which makes it difficult for them to engage in prison rehabilitation programmes (Ward 2007; Leith 2011). Low self-esteem, poor impulse control and related behavioural problems, as well as a limited ability to resolve conflicts constructively, are prominent examples of mental health-related problems common among prisoners. In addition, low empathy is strongly related to aggressive and offending behaviour (Jolliffe 2004). Mental health problems are a significant issue for a majority of offenders and therefore something that needs to be addressed alongside (or in some instances instead of) offending behaviour (Priebe 2008).

Psychological characteristics of offenders vary according to age, gender and crime type. Young offenders are more likely than adults to have mental health problems and suicidal tendencies (Prison Reform Trust 2011). Female offenders tend to have more mental health problems than male offenders and are also more likely to have experienced emotional, physical, sexual or financial abuse (or a combination) (James 2006). Juvenile offenders tend to have attention and hyperactivity problems, academic difficulties, oppositional behaviour, peer relationship and social skills deficits, cognitive and attributional deficiencies, anger management problems and impulsivity (Wyatt 2002). In terms of crime type, sexual offenders seem to be more likely to have poor social skills and low self-esteem than other offenders (Worling 2001).

**Description of the Intervention**

Music therapy is commonly defined as a "systematic process of intervention wherein the therapist helps the client to improve health, using music experiences and the relationships that develop through them as dynamic forces of change" (Bruscia 1998). This definition encompasses all theoretical models of music therapy and includes both treatment and prevention/health promotion. However, it does not include music interventions at an 'auxiliary level' (Bruscia 1998), such as music listening without the presence of a therapist; such interventions are sometimes referred to as music medicine (Gold 2009; Gold 2011a). Music therapists are specifically trained to intervene within the music (Gold 2007), and it is therefore seen as important that the intervention is administered by an appropriately credentialed music therapist to ensure the quality of the intervention.

The music experiences offered in music therapy often include active music making, either as spontaneous creative expression (for example, free or structured improvisation; Albornoz 2011), as a compositional process (for example, songwriting; Edgerton 1990; Baker 2011) or as reproduction of existing musical material (for example, singing; Clark 2012) in individual or group settings. However, listening to music can also be a central modality and often serves as an anchor point for verbal reflection (Blom 2011). The method of working, as well as the level of structure and the degree of emphasis on the music itself versus discussion of personal issues, may vary. These are typically adapted in response to the client's needs and wishes, but have also been an issue of debate among music therapists (Mössler 2011a). Music therapy has a long history of application in mental health (Gold 2009; Mössler 2011b) and also in correctional institutions, where its use dates back to at least the 1930s (Codding 2002). Various authors have described the use of music therapy methods such as improvisation (Hoskyns 1988), songs and metaphoric imagery (Chambers 2008), creating and performing music (O'Grady 2009) and music relaxation (Thaut 1989a) in music therapy with offenders, either in individual or small group settings (Rio 2002).

Goals of music therapy with offenders are usually broad. According to a survey among 49 American music therapists, the following are typical: to provide "a non-threatening, motivating reality focus for use of leisure time and release of energy" (100%), to promote "self-esteem" (94%), "self-control" (91%), "appropriate release of tension, stress and anxiety" and "coping skills" (91%); 32 further goals were listed by Codding 2002. Some researchers have suggested that goals of interventions for offenders should be categorised according to whether or not they are risk or protective factors for future criminal behaviour (Bonta 2007). Examples of risk factors are antisocial personality patterns, pro-criminal attitudes, substance abuse, poor family relationships, poor behavioural control; examples of factors that help people to move away from crime (protective factors) are positive social orientation, intolerant attitude towards deviance, a flexible personality, good coping skills and self-calming ability (Bonta 2007). A dynamic risk factor is a risk factor that can change. Negative emotions and thoughts, low coping skills, lack of anger management ability, lack of empathy and impulsivity may be examples of dynamic risk factors. A qualitative study that collected the views of creative arts therapists working with offenders found music therapy to be focusing on positive developmental goals such as coping skills and emotion regulation (Smeijsters 2011).

However, the link between mental health issues and criminal risk is not always entirely clear. There are also numerous reports of music therapy focusing on other goals that are relevant for the prisoner's health but may or may not be related to the risk of future criminal behaviour, such as emotional expression and assertive behaviour (Cohen 1987), mood, relaxation and insight (Thaut 1989a) or decision-making style (Moss 2004). A certain tension between goals related to mental health and goals related to criminal risk may be due to the double nature of therapies for offenders. Many studies of music therapy for offenders explicitly targeted a population who also had a mental health problem (for example, Thaut 1989a; Codding 2002); even where this was not the case, difficult life histories and traumatic experiences were almost ubiquitous and an important focus of therapeutic work (for example, O'Grady 2009).

**How the Intervention Might Work**

Music therapy is a complex intervention. Explaining mechanisms of change in complex interventions is often not straightforward because of multiple components interacting with each other, and because the context influences the outcome (Craig 2008, Gold 2012). Although a model has been suggested for explaining change in externalising behaviour problems through cognitive-behavioural music therapy (Hakvoort 2013), a more general model of change for music therapy with offenders has not been proposed previously.

Figure 1 shows an attempt to explain the pathways in which music therapy might work for offenders. Starting from the definition cited above (Bruscia 1998), the two basic aspects of music therapy are interpersonal interaction and music experiences (first column in the figure). Both aspects are connected to each other and cannot easily be separated, although the extent to which either the music or the interaction aspects is the main agent might vary. Basic guiding principles (above the first column) include analogy, metaphor, and aesthetics. Analogy and metaphor are two closely related concepts that basically state that the way people express themselves in music is related to the way they act in other situations. Thus, the way people act in general, or the repertoire they have for interacting with others, may be changed or expanded by changing the way they interact in music (Smeijsters 2012). Analogy is a more elementary concept than metaphor and is related to very early relational learning (Stern 2010). In addition, aesthetic experience (where 'aesthetic' is understood much more broadly than to be 'pleasing' in a narrow sense) is seen as a basic part of the human condition (Aigen 2007). This might explain why music therapy is often found to be motivating for clients who are not easily motivated for other therapies. Motivation is therefore an overarching principle towards achieving direct outcomes (second column). Interpersonal interaction through music has -by means of the basic principles described -a direct and immediate quality that lends itself well to learning very basic communicative abilities (Stern 2010), to experience and be able to give social support (Procter 2011), and to improve the ability to understand and share the feelings of others, i.e. empathy. It is proposed that improvement in mental health outcomes (third column) occurs primarily through those direct outcomes. Both externalising (e.g., aggressive or impulsive behaviour) and internalising problems (e.g. anxiety, depression) may be influenced. These are separate but connected (i.e. correlated and interacting with each other). Finally, recidivism and quality of life are downstream outcomes that might be influenced indirectly by means of reduced externalising and internalising mental health problems.

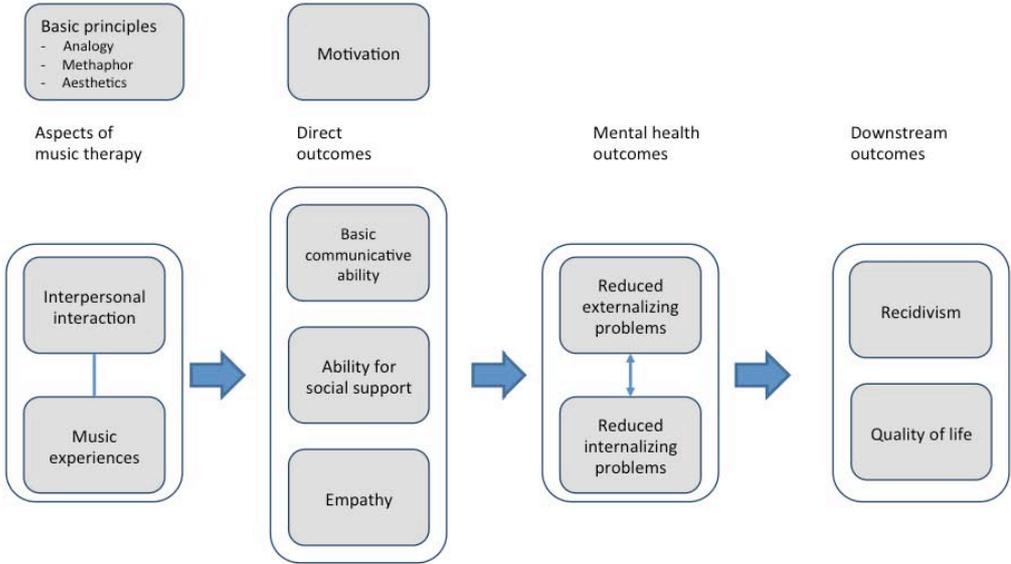

Figure 1. Conceptual model of music therapy for offenders

In summary, using music as a medium in therapy might be beneficial because:

Music is a "language of emotions" (Juslin 2010) and may help people develop the ability to perceive, express and change emotions (e.g. Erkkilä 2011; Mössler 2011b; Gold 2013). For offenders who have difficulty experiencing and identifying emotions, not necessarily with a diagnosed disorder, music therapy can offer a safe and contained space in which the client is supported in identifying and exploring emotional states as well as learning to express and regulate them in a positive way (Loth 1994; Loth 1996). Strong and potentially destructive emotions that might otherwise be expressed through self-harming or challenging behaviour can find positive and appropriate expression in music (Leith 2011).

Music-making is also highly social in nature (Stern 2010, Procter 2011). Making music together may therefore be an effective way of improving the ability to build and sustain social relationships, to communicate with and to relate to others. Because music is social, music therapy may enhance offenders' listening and sharing skills. It may foster flexibility and sensitivity to others within the musical relationship with the therapist and, in group therapy, with other group members. These interpersonal relationships, even attachments developed in music experiences, have significant therapeutic meanings, especially for those who were lacking these experiences in the past (Rio 2002). It also creates a bridge for the offenders to link their life inside of prison and outside realities (O'Grady 2011, Tuastad 2013). It may strengthen an awareness of self in relation to others and a client's confidence in their own ability to build relationships and to find positive ways of making their needs known (Leith 2011).

Music is a non-verbal medium. It conveys meanings, but not in the same way as verbal language. This makes it motivating for clients who are unmotivated for verbal therapies (Gold 2013). In a setting where verbal self-disclosure can be detrimental for the individual, purely verbal therapy may be of limited value. Music therapy may be valuable for clients who respond best to action-oriented forms of expression (Nolan 1983). It may create a healthy and non-threatening environment in which offenders can overcome anxiety and resistance, increase motivational ties to reality and gain insight through a learning process that is perceived as meaningful. Music also provides multimodal experiences, such as feelings, internal images and body sensations (Skaggs 1997).

The current literature does not suggest any differences in mechanisms of change between age groups, genders, or types of offences. Therefore it appears justified to combine them in one review. If future research should suggest substantial differences between those groups, this will be reconsidered.

**Why It Is Important to Do this Review**

Both the application of and research on music therapy for offenders seem to be increasing. Numerous case reports and some outcome studies have suggested beneficial effects, but no systematic review of effectiveness has been conducted yet. The present review is therefore necessary to inform policy-makers and practitioners, as well as to guide future research in the field.

**OBJECTIVES**

To assess the effectiveness of music therapy on adolescent and adult offenders in custodial institutions including forensic psychiatric hospitals, and offenders or probationers in the community. The outcomes to be evaluated include alleviated symptoms of mental illness, psychosocial competencies and reduced recidivism.

**METHODS**

**Criteria for Considering Studies for this Review**

**Types of studies**

Randomised controlled trials (RCTs), including quasi-randomised controlled trials (for example, allocation by the person's date of birth, by the day of the week or month of the year, by a person's medical record number or alternate allocation). Studies will be considered regardless of sample size. It is likely that any existing studies in this field will be small and underpowered pilot studies, but such studies may still contain important information for building larger studies in the future (Gold 2004b). The limited power of small studies will be considered in the interpretation.

**Types of participants**

Adult and juvenile offenders irrespective of offence, gender or nationality, with or without formally diagnosed mental illness, in any institutional setting (for example, forensic psychiatry, prison, a correctional institution) or in the community. The starting age for criminal responsibility varies considerably across countries (Prison Reform Trust 2011, p. 34), but is 14 in many countries (Hazel 2008). We therefore aim to include people from 14 years upwards if in a community setting. Participants in institutional settings will be included if the institution serves adult or juvenile offenders as defined in the place where the study was conducted.

**Types of interventions**

1. Music therapy (either alone or in addition to a standard care that would also be provided in the comparison condition). Music therapy is defined as "a systematic process of interventions wherein the therapist helps the client to promote health, using musical experiences and the relationships that develop through them as dynamic forces of change" (Bruscia 1998). The methods of music therapy include improvisation, re-creating, composing, listening and variations of them (Bruscia 1998). Music therapy as offered in the trials must conform to this definition; in particular it must include music experiences as well as a relationship with an appropriately credentialed music therapist.

2. Standard care, no treatment or any kind of 'placebo' therapy. The term placebo therapy

does not refer to a medication. Instead, it is a kind of pseudo therapy aimed at controlling for the therapy's unspecific effects, for example by constructing an intervention that is similar to the active interventions but supposedly does not contain its active ingredients. Examples in music therapy can be found (Gold 2006) but their relevance is controversial (Gold 2011b). See also Wampold (2001) for a further discussion of 'placebo' therapies.

**Types of outcome measures**

General considerations

All primary outcomes are marked with an asterisk (*), indicating that we will include them in a 'Summary of findings' table. If there are several time points for the same outcome from the same study, we will categorise them as short term (up to 12 weeks), medium term (13 to 26 weeks) or long term (more than 26 weeks after randomisation), as in a previous review (Mössler 2011b). When outcomes are reported on a scale, they should either be based on an independent observer (that is, not the music therapist) or on self-reports. Scales should preferably be standardised and published with known reliability and validity.

Primary outcomes

Primary outcomes are those that have the greatest importance. They must be relevant for the participants and related to the intervention's goals. The following primary outcomes are based on the most common goals of music therapy for offenders (Codding 2002; Smeijsters 2011) as well as on the proposed model of change (Figure 1 and How the intervention might work).

1. Self-concept/self-esteem/self-efficacy *

Almost all music therapists working in correctional settings (94%, Codding 2002) endorse self-esteem as an important goal. Related constructs are self-concept and self-efficacy. The three constructs are often used interchangeably and are therefore considered together in this review. Self-esteem emphasises more the affective components of judging one's self-worth, whereas self-concept and self-efficacy emphasize more the cognitive aspects. Sources disagree on whether self-esteem is a part of self-concept or vice versa (Judge 2001). Low self-esteem may contribute to externalising behaviour problems and delinquency, but also to internalising problems such as depression (Baumeister 2003). However, self-efficacy might be a clearer predictor for behavioural outcomes than self-esteem (Bandura 1997). This outcome is related to participants' sense of identity, which is often impaired or not fully developed. We consider it therefore as a relevant outcome in its own right, not only as a predictor of downstream outcomes. However, ceasing to see oneself as an offender and developing a more positive identity may also be linked to desistance from crime, especially for people with deeply entrenched identities as offenders (Maruna 2003). This outcome is subjective in nature and will normally be measured on a self-report scale as a continuous variable.

2. Behaviour management *

This includes 'self-control' and appropriate management of aggressive or self-harming behaviour and impulsivity. This outcome was described as important by 91% of music therapists (Codding 2002), and some music therapy approaches focus primarily on this outcome (Hakvoort 2013). It is an objective, observable outcome of immediate relevance to prison management and very likely relevant for criminal risk, as the offence itself is a deviant behaviour. It may be measured as continuous or count data, based on independent observer reports or possibly through self-reports.

3. Anxiety *

Among mental health-related outcomes, anxiety seems to be an important. "Appropriate release of tension, stress and anxiety" has been endorsed as a relevant goal by 91% of music therapists (Codding 2002). Along with depression, it is one of the most important internalising behaviour problems. Both anxiety and depression might be considered as primary outcomes because they are closely related. However, given that the number of primary outcomes should be limited, we made a choice to prioritize anxiety as it seems to be more often endorsed as an outcome and possibly also more prevalent among offenders.

4. Empathy *

Empathy is defined as "the ability to understand and share the feelings of another" (Oxford 2010). It is a direct outcome of music therapy because to work with music experiences is to work with the ability to understand and share feelings. Many offenders, specifically those with antisocial personality traits, have insufficiently developed empathy. To "develop trust and empathy" is a relevant goal for 84% of music therapists (Codding 2002).

5. Any adverse event *

Little is known about the potential adverse effects of music therapy (Edwards 2011). No specific adverse effects of music therapy have been described in the literature. Therefore, any adverse events will be reported here.

Secondary outcomes

Secondary outcomes concern additional effects of music therapy that may be relevant only for some offenders or that are relevant but not directly targeted by music therapy.

1. Depression

Many offenders suffer from negative mood states other than anxiety, such as symptoms of depression. "To reduce the number and length of depressive episodes" was endorsed as a relevant goal by 56% of music therapists (Codding 2002). Music therapy may be well-suited to address depressive symptoms (Erkkilä 2011). Together with anxiety, it is an important internalising problem.

2. Quality of life

Quality of life is a general goal that applies to almost any health intervention for most conditions. Improved quality of life is also likely associated with reduced criminal risk (Ward 2004). Many aspects of offenders' quality of life may be beyond the areas that music therapy can address, and it does not seem to be among the most common goals of music therapists in the field (Codding 2002). It is however a downstream outcome that may result from improved mental health. In some contexts it may be controversial whether quality of life of offenders should be improved at all.

3. Substance abuse

Substance abuse is an internalising mental health problem. It is often considered together with anxiety and depression (Albornoz 2011). Many offenders suffer from it, and it also contributes to offending behaviour.

4. Recidivism

To reduce future criminal activity should be an ultimate goal of any intervention for offenders, including music therapy. However, this outcome will also be influenced by many external factors. Therefore recidivism is considered as a downstream effect, by means of direct and mental health outcomes (as shown in Figure 1), rather than as a direct outcome. Behaviour management (listed under primary outcomes) may be considered an early indicator of this outcome.

**Search Methods for Identification of Studies**

**Electronic searches**

We will search the following databases, with no date or language restrictions: Cochrane

Central Register of Controlled Trials (CENTRAL), part of The Cochrane Library; MEDLINE; EMBASE; PsycINFO; LILACS; CINAHL; ERIC; Sociological Abstracts; International Bibliography of Social Sciences; National Criminal Justice Reference Service Abstracts; RILM Abstracts of Music Literature; Social Science Citation Index; SCOPUS; Conference Proceedings Citation Index -Social Science & Humanitee; WorldCat (theses search); Rutgers School of Law Grey Literature Database; ClinicalTrials.gov; ICTRP; *meta*Register of Controlled Trials.

We will search MEDLINE using the following search strategy, which will be adapted for the other databases using appropriate syntax and controlled vocabulary. The strategy will be simplified for databases and websites that do not support complex search strings. Language and date limits will not be applied and we will not use a study methods filter. We will seek translations of papers when necessary. 1. music therapy/ 2. music$.tw. 3. (guided imagery adj3 music).tw. 4. BMGIM or GIM.tw. 5. (vibro-acoustic$ or vibroacoustic$).tw. 6. music/ 7. (sing or singing or song$ or choral$ or choir$).tw. 8. (percussion$ or rhythm$ or tempo).tw. 9.melod$.tw. 10. improvis$.tw. 11. (Nordoff-Robbin$ or Bonny$).tw. 12. ((auditory or acoustic or sound$) adj5 (stimulat$ or cue$)).tw. 13. ((play$ or learn$) adj3 instrument$).tw. 14. or/1-13 15. prisoners/ 16. prison/ 17. juvenile delinquency/ 18. (borstal$ or convict$ or correctional$ or criminal$ or custody OR custodial OR delinquen$ or detain$ or detention$ or gaol$ or imprison$ or incarcerat$ or inmate$ or in-mate$ OR jail$ or offenc$ OR offens$ OR offender$ or penal$ or penitentiar$ or prison$ or probation$ or recidivid$ OR reformatory or (reform adj school$) or (secure adj accommodation)).tw. 19. residential treatment/ 20. psychiatric hospitals/ 21. (forensic adj3 (hospital$ or patient$)).tw. 22. or/15-21 23 14 and 22.

**Searching other resources**

Grey literature

We will search the websites of relevant professional and research organisations to retrieve grey literature, including the World Federation of Music Therapy, the American Music Therapy Association, the European Music Therapy Confederation, the Chinese Music Therapy Association, the Hong Kong Music Therapy Association, the Japanese Music Therapy Association, the Korean Music Therapy Association, the Canadian Association for Music Therapy, the Australian Music Therapy Association, the Music Therapy Association of Taiwan, the British Association for Music Therapy and Music in Prisons

(www.musicinprisons.org.uk). We will also search the criminal justice grey literature database of Rutgers School of Law (law-library.rutgers.edu/cj/gray/).

Handsearching

We regularly monitor the content of following journals and will continue to do so (XJC: all issues of the Journal of Music Therapy since the journal started in 1964; HL: British Journal of Music Therapy since 1980; Australian Journal of Music Therapy since 2006; CG: Nordic Journal of Music Therapy since the journal started in 1992, Musiktherapeutische Umschau since 1998). Additionally, extensive handsearching of music therapy journals was performed earlier for related reviews (Gold 2004a; Maratos 2008; Gold 2009; Mössler 2011b) and any relevant records retrieved there will be re-used.

Correspondence

We will attempt to contact the authors of relevant studies for additional or missed studies and for further resources and information.

Reference lists

Any potentially relevant studies in the reference lists of studies retrieved by the searches will be followed up.

**Data Collection and Analysis**

**Selection of studies**

Duplicate records of the same report will be removed using reference management software. Two review authors (XJC and HL) will independently examine titles and abstracts and exclude those reports that are clearly not randomised studies, not about offenders or not about music therapy. Two review authors (XJC and HL) will then independently examine the full text of potentially relevant reports and decide which studies meet the eligibility criteria. If disagreement occurs, all three review authors will discuss to reach a resolution. Where information on an eligible study is missing, we will attempt to contact the trial authors to obtain that information.

**Data extraction and management**

Two review authors (XJC and HL) will independently extract data for each trial using a data extraction form to collect information about the population, interventions, randomisation method, blinding, sample size, outcome measures, follow-up duration, attrition and handling of missing data, and methods of analysis.

**Assessment of risk of bias in included studies**

We will use The Cochrane Collaboration's tool for assessing risk of bias (Higgins 2011). Two review authors (XJC and HL) will independently assess the risk of bias within each included study based on the following domains, with a review author's judgments presented as answers of 'high', 'low' or 'unclear' risk of bias. Any disagreements will be resolved by discussion and disagreements will be arbitrated by a third review author (CG). The tool will be used to assess the following domains: randomisation, allocation concealment, blinding of outcome assessment, incomplete outcome data, and selective outcome reporting and other sources of bias.

**Randomisation**

Randomisation will receive the following judgments. 'Low' when participants were allocated through a truly randomised sequence (such as computer-generated random numbers, a random numbers table, coin-tossing). 'Unclear' when the randomisation method was not clearly stated or unknown. 'High' when randomisation did not use an appropriate method of sequence generation.

**Allocation concealment**

'Low' when allocation concealment was clearly stated in the study. 'Unclear' when allocation concealment was not clearly stated or unknown. 'High' when allocation was not concealed from either participants or researchers before informed consent, or from researchers before decisions about inclusion were made.

**Blinding**

a) Blinding of assessors. For all outcomes that are not self-reports, we will examine whether blinding of outcome assessors was attempted, and if and how success of blinding was verified. Quality of blinding will receive the following judgments. 'Low' when assessors were blind to the treatment conditions. 'Unclear' when blinding of assessors was not reported. 'High' when assessors were not blind to treatment conditions.

b) Blinding of therapists and clients. We will also consider whether therapists and clients were blinded to the intervention, even though we do not know of any method to ensure blinding of therapists and clients in a psychosocial intervention (Gold 2011c).

**Addressing incomplete outcome data**

Assessment will take into account whether researchers used intention-to-treat analyses by including measurements from all the participants, including those who did not participate fully in the treatment protocol. Those studies where the researchers did not use intention-to-treat analyses and it is not possible to conduct them with the available data will be identified. The adequacy of the way the authors of the trials dealt with missing data will receive the following judgments. 'Low' when intention-to-treat analyses were used or can be performed using available data. 'Unclear' when information about whether intention-to-treat analyses were performed was not available and cannot be acquired by contacting the researchers of the study. 'High' when intention-to-treat analyses were not performed and cannot be done using available data. Balance of drop-outs and reasons for dropping out will be explored. We will use sensitivity analysis to assess the impact of drop-outs.

**Selective reporting**

The likelihood that the authors of the trial omitted some of the collected data when presenting the results will be determined and will receive the following judgments. 'Low' when all collected data seem to be reported. 'Unclear' when it is not clear whether other data were collected and not reported. 'High' when the data from some measures used in the trial are not reported.

**Other bias**

Assessment will determine whether any other bias is present in the trial. In particular, appropriate administration of the intervention (adequacy of music therapy method and training) will be assessed as an important issue that may affect the results (Mössler 2011b). Inappropriate administration of an intervention is one of the examples of potential biases listed in the Cochrane Handbook for Systematic Reviews of Interventions (Higgins 2011, Section 8.15.1.5). Baseline imbalance will also be assessed as a potential source of bias.

**Measures of Treatment Effect**

**Dichotomous data**

For dichotomous data, we will calculate an odds ratio (OR) with a 95% confidence interval (CI) for each outcome in each trial (Higgins 2011). From our knowledge of the field, risk ratios (RR) are rarely reported and if so, they are reported with the original cross table data, making it possible to calculate any measure of association. Risk hazards are even more uncommon. If any such measures should occur we will analyse them in line with the Cochrane Handbook for Systematic Reviews of Interventions (Higgins 2011).

**Continuous data**

When means and standard deviations (SD) are provided in the study, we will analyse continuous data. Effect sizes (that is, standardised mean differences (SMD)) will be used because they facilitate clinical interpretation (Cohen 1988; Gold 2004b) and because they allow for combining results from different scales for the same outcome. (The type of effect size used will be Hedges' g, which is similar to Cohen's d but with a small-sample bias correction.)

**Unit of Analysis Issues**

**Cluster randomisation**

If cluster-randomised trials are included, we will examine whether the clustering was taken into account in the analysis. If clustering has not been taken into account in the published analysis, we will attempt to adjust the analysis using the design effect. For this adjustment, we will ideally use an estimate of intra-class correlation from the same study, but where this is not available, we will attempt to find and use an external estimate from a similar study (Ukoumunne 1999).

**Studies with multiple intervention groups**

For studies where there are multiple intervention (or control) groups, we will merge groups (if deemed similar enough) to avoid multiple comparisons. The data from the same group will not be analysed twice in the same meta-analysis. A separate meta-analysis will be done for each comparison.

**Multiple time points**

If outcomes were measured more than once in the same study, we will not combine different time points in the same meta-analysis, but will use different meta-analyses to avoid unit of analysis issues. If studies used an appropriate analysis that takes the dependence of longitudinal data into account (for example, generalised estimating equations or linear mixed models), such results will be reported narratively.

**Multiple measures of the same outcome**

If more than one measure was used to assess the same outcome, and these measures were interchangeable (for example, both were standardised scales, had the same level of blinding and were used at the same time point), we will only use the measure that was identified as primary in the original study. Where this is not possible, we will attempt to make an informed judgment as to which measure was likely to be intended as primary.

**Dealing with missing data**

We will assess missing data in the included studies. The meta-analysis will be based on the data of all original participants when it is possible. If a study reports outcomes only for participants completing the trial, or only for participants who followed the protocol, efforts will be made to contact study authors to request missing data. We will report the reasons and types of missing data. If it is not possible to conduct an intention-to-treat analysis in some trials, we will use sensitivity analysis to assess the potential bias introduced by those trials.

**Assessment of heterogeneity**

We will assess heterogeneity in the following areas: clinical diversity, methodological diversity and statistical heterogeneity. We will assess the first two types of heterogeneity prior to combining trials in a meta-analysis. We will assess statistical heterogeneity in a meta-analysis using the $I^2$ statistic, a descriptive measure that represents the percentage of variability that is due to heterogeneity rather than sampling error or chance. The percentages obtained using the $I^2$ statistic will be interpreted according to the Cochrane Handbook for Systematic Reviews of Interventions (Higgins 2011, Section 9.5.2). The $Chi^2$ test will be used in addition to assess the presence of heterogeneity. Because this test is often underpowered when there are few studies, a P value smaller than 0.10 will be interpreted as an indication of possible heterogeneity of intervention effects. We will conduct subgroup analyses to explore the sources of heterogeneity. We have not defined a fixed cut-off value for $I^2$. As described in the Cochrane Handbook for Systematic Reviews of Interventions (Higgins 2011, Section 9.5.2, p. 278), there cannot be one percentage at which heterogeneity is "too high". Rather, the importance of the observed value depends on the magnitude and direction of effects as well as the strength of evidence for heterogeneity (P value or CI).

**Assessment of reporting biases**

For avoiding publication bias, we will make efforts to obtain and include data from unpublished trials that meet the inclusion criteria. We will use a funnel plot to assess the likelihood of publication bias in meta-analyses with enough trials. We will include a triangular 95% confidence region based on a fixed-effect meta-analysis in the funnel plot, and different plotting symbols will be used to identify different subgroups if applicable. Funnel plots will be interpreted cautiously as recommended in the Cochrane Handbook for Systematic Reviews of Interventions (Higgins 2011, section 10.4.5), and if asymmetries are found we will examine clinical variation in the studies.

**Data synthesis**

We will conduct a meta-analysis to combine data from studies using the same comparisons and outcomes. Different comparisons and outcomes will be meta-analysed separately. There are divergent views on the question of random versus fixed effects. Although random effects are often seen as an appropriate solution when there is unexplained heterogeneity, the Cochrane Handbook cautions that "a random-effects model does not 'take account' of the heterogeneity in the sense that it is no longer an issue" (Higgins 2011, 9.5.4, p. 280). Some have even argued that "combining heterogeneous studies using the random-effects model is a mistake and leads to inconclusive meta-analyses" (Al Khalaf 2011). Fixed-effects models are also more straightforward to interpret (Higgins 2011). However, it is not possible to resolve this controversy as part of this review. We will therefore follow the advice given by the editor and the statistical reviewer: We will, by default, apply both fixed and random effects, and if the 95% CI of the random-effects analysis includes the 95% CI of the fixed-effect analysis, report only the random-effect analysis.

**Subgroup analysis and investigation of heterogeneity**

We aim to limit subgroup analyses to the most important ones. We have selected age and gender, but others (e.g. age of onset of offending behaviour) may also be relevant. Should we find heterogeneity we will examine the following subgroups.

1. Young offenders versus adult offenders. Young offenders are an important subgroup because they are more likely to have mental health problems. Although varying cut-offs have been used in the literature, it seems to be most common to regard those under 21 years of age as young offenders (Prison Reform Trust 2011, pp. 37-38; a cut-off of 25 years has also been used in the literature) and we therefore aim to use this cut-off.

2. Male versus female offenders. Women are more likely than men to have mental health problems, which often find expression in self-harming behaviour (James 2006). Women prisoners in the UK commit around 50% of self-harm incidents although they represent only 5% of the total prison population (Smee 2009). The level of importance and nature of association of criminogenic factors may also differ between genders (Blanchette 2001).

**Sensitivity analysis**

We will use a sensitivity analysis to investigate if trial characteristics such as unclear methodology or missing data may have influenced the results of the analyses, by repeating the analysis with the problematic studies removed. In particular, we will use sensitivity analyses to examine the influence of overall risk of bias (excluding studies with a high risk of bias) and quality of the intervention (excluding studies that relied on inappropriate music therapy methods or training level). Randomisation method and other bias (quality of the intervention) are anticipated to be the most important sources of bias in this area.

**ACKNOWLEDGEMENTS**

James Tyler Carpenter and Laurien Hakvoort provided valuable feedback on an earlier version of this protocol.

**CONTRIBUTIONS OF AUTHORS**

XJC conceived the review, drafted the first version of the protocol and suggested selection criteria. HL and CG helped with drafting and re-writing the protocol, provided feedback on selection criteria and other methods details. LEA and TTM helped with improving the sections on how the intervention might work and description of outcomes.

**DECLARATIONS OF INTEREST**

XJC, HL and CG are trained music therapists. XJC and CG are involved in a potentially eligible ongoing study (ClinicalTrials.gov Identifier: NCT01633125). CG is also involved in another potentially eligible study (ISRCTN22518605) and serves as an associate editor of the Cochrane Developmental, Psychosocial and Learning Problems Group. HL works as a music therapist with women in prison and is conducting mixed methods research into music therapy and the resettlement of women prisoners with non-psychotic mental health problems. LEA and TTM declare that they have no conflict of interest.

**SOURCES OF SUPPORT**

**Internal sources**

PhD Programme in Music Therapy, Aalborg University, Denmark.

**External sources**

GC Rieber Foundation, Bergen, Norway.

The Research Council of Norway, Norway.27